\documentstyle[aps,preprint,epsfig]{revtex}

\def \beq {\begin{equation}}
\def \eeq {\end{equation}}

\begin{document}

\draft

\title{Quantum effects and superquintessence \\ in the
new age of precision cosmology}

\author{
E. Gunzig$^{1}$,
Alberto Saa$^{2}$ }

\address{1)
RggR, Universit\'e Libre de Bruxelles, 
CP 231, 1050 Bruxelles, Belgium.
}

\address{2)
IMECC -- UNICAMP,
C.P. 6065, 13081-970 Campinas, SP, Brazil.}

\maketitle

\begin{abstract}
Recent observations of Type Ia supernova at high redshifts
establish that the dark energy component of the universe
has (a probably constant) ratio between pressure and
energy density $w=p/\rho=-1.02\left(^{+0.13}_{-0.19}\right)$.
The conventional quintessence models for dark energy are
restricted to the range $-1\le w < 0$, with the cosmological
constant corresponding to $w=-1$. Conformally coupled quintessence
models are the simplest ones compatible with the marginally 
allowed superaccelerated regime ($w<-1$). However, they are
known to be plagued with anisotropic singularities.

We argue here that the extension of the classical approach
to the semiclassical one, with the
inclusion of quantum counterterms 
necessary to ensure the renormalization, 
can eliminate the anisotropic singularities preserving
the isotropic behavior of conformally coupled superquintessence
models. Hence, besides of having other interesting properties,
they are consistent candidates to describe the 
superaccelerated phases of the universe compatible with
the present experimental data.

\end{abstract}
\pacs{ }

We are entering a new era in cosmology. New generations of
experiments are providing us with wealthy sets of experimental
data, with a precision simply unbelievable in a very recent past,
no more than 20 years ago. New observations of Type Ia Supernova,
done from the Hubble Space Telescope (see, for instance, \cite{hubble})
and from earth telescopes\cite{CFH} are improving considerably
the experimental evidences that our universe is undergoing a
phase of accelerated expansion, as first noticed in\cite{perlmutter}.
Einstein's equations require a negative pressure cosmological fluid 
acting as source
to an accelerated expansion phase of the universe. Such repulsive 
fluid must be nonluminous,
otherwise it would be detected. This is the ``dark'' energy component that
dominates the evolution of the universe today. Moreover,
it corresponds to 70\% of the composition of the universe.
Steinhardt\cite{steinhardt1} considers the discovery of dark
energy as one of the most surprising and profound of the
history of science, and according to him, ``{\em we are probably the
last generation to have been taught that gravity aways
attracts}''.

The observations of the fluctuations in the cosmic microwave background
(CMB),
mainly the spectacular results of WMAP\cite{wmap}, have enforced
the dark energy hypothesis. Despite the severe 
degeneracy problems in the CMB parameters analysis,
the existence of a dark energy component represented by
a small cosmological constant $\Lambda$ is so favored 
by the available data
that the standard cosmological model is now called 
the $\Lambda$CDM\footnote{CDM stands to cold dark matter, a nonluminous
gravitationally attractive 
form of matter, responsible by approximately 25\% of the energy
content of the universe.} model.

An important feature of the dark energy fluid is the ratio $w$
between its pressure and energy density. For the cosmological
constant, $w=-1$. Dark energy can also be described by a
field\cite{caldwell}, the so-called quintessence:
a dynamical field with negative pressure, not necessarily
homogeneously distributed, in contrast to the cosmological constant.
Typically,   quintessence models are constructed 
from the Einstein-Hilbert action with
minimally coupled scalar fields:
\beq
S=\int d^4x \sqrt{-g}\left\{R - \partial_a\phi\partial^a\phi
-2V(\phi) \right\}.
\eeq
For such models, the ratio between the pressure and the
energy density of a homogeneous scalar field $\phi$
\beq
w = \frac{p}{\rho} = \frac{\dot{\phi}^2-2V(\phi)}{\dot{\phi}^2+2V(\phi)}
\eeq
is bounded from below, $w\ge -1$. According to the
potential $V(\phi)$, $w$ can vary between -1 and 0. Typically, $w$ depends
on time in quintessential models. 

The recent results of \cite{hubble},
obtained from Type Ia supernova at $z>1$, 
impose strong constraints on rapidly evolving quintessential
models, favoring scenarios with constant $w$. 
According to the experimental data, there are   some marginal evidences suggesting
that $w<-1$. Indeed, according to \cite{hubble}, 
$w=-1.02\left(^{+0.13}_{-0.19}\right)$. If one really has $w<-1$, both
cosmological constant and minimally coupled scalar quintessence
descriptions for dark energy are ruled out.
A regime with $w<-1$ is called superaccelerated, and a model
achieving it, superquintessence\cite{faraoni}. The simplest
non-exotic model exhibiting superaccelerated expansion is
that of a scalar field conformally coupled with quartic
potential\cite{PRD}, corresponding to the action
\beq
\label{act}
S=\int d^4x \sqrt{-g}\left\{F(\phi)R - \partial_a\phi\partial^a\phi
-2V(\phi) \right\},
\eeq
with $F(\phi)=1-\frac{1}{6}\phi^2$
and $V(\phi)=\frac{m}{2}\phi^2-\frac{\Omega}{4}\phi^4$.
For the homogeneous and isotropic case,
{\bf all} the solutions of (\ref{act}) are
regular ones, and some of them presents  
other interesting
dynamical behaviors besides the  
superaccelerated regime, as the possible 
avoidance of big-bang singularities\cite{PRD}.

However, the hypersurface $F(\phi)=0$, the frontier 
between the
regions where gravity is effectively attractive ($F>0$) and
repulsive ($F<0$) is known to be problematic.
Starobinski was the first to notice
it in a particular anisotropic model\cite{Starobinski}. 
The standard perturbation theory for helicity-2 and helicity-0
excitations,
derived directly from the action (\ref{act})
fails on $F(\phi)=0$\cite{G}. One can show, indeed, that
the hypersurface $F(\phi)=0$ corresponds generically to
an unavoidable anisotropic singularity\cite{PRD2}. 
Any anisotropic
solution crossing it will end in a spacetime singularity,
no matter how small is the anisotropy. Such an unstable behavior is,
of course, unacceptable for any cosmological model. 
Unfortunately,
all the interesting solutions of the model presented in \cite{PRD}
cross the hypersurface $F(\phi)=0$.

It is quite easy to understand the origin of the 
anisotropic singularities
studied in \cite{PRD2} for the models (\ref{act}) with general
$F(\phi)$. As the kinetic term for the metric $g_{ab}$ is
$F(\phi)R$, we have no guarantee that the Cauchy problem is
well posed on the hypersurface $F(\phi)=0$. Indeed, for the anisotropic
case, it is not, meaning that one cannot assign
any value of $\dot{\phi}$, $g_{ab}$ and $\dot{g}_{ab}$
when $F(\phi)=0$. On the contrary, the Cauchy problem
is well posed everywhere in the isotropic case. This is the
reason why all the homogeneous solutions of (\ref{act}) studied
in \cite{PRD} are regular.

The relevant question arising here is if one should give up
of conformal coupling to construct superquintessence models
due to their anisotropic instabilities or not. Conformally coupled
superquintessence models are the simplest ones 
that we can construct with 
homogeneous and isotropic solutions compatible with 
  superaccelerated phases of the universe. 
They also have other particular
predictions that could be checked experimentally
\cite{faraoni}.
Our
question is obviously related to another one: is it possible
to modify the model (\ref{act}) in order to avoid the 
anisotropic singularities and, at the same time, preserve
its isotropic solutions? The answer is, fortunately, affirmative.
Moreover, the desired modification is achieved by the inclusion of
a new term coming from semiclassical analysis
\footnote{By semiclassical here one means that $\phi$ is quantized on a
classical gravitational background.}.

The idea of incorporating vacuum semiclassical 
effects into gravity has a long
history, and a good set of references is presented in
\cite{Birrel,Buchbinder}.
Zeldovich was the first to propose\cite{Zeldovich}, in 1967, 
that a cosmological
constant term could arise from quantum considerations of matter.
Yet in the sixties, in a set of seminal works, Parker 
considered\cite{Parker} 
the effect of the creation
of particles in a expanding universe, and discussed the possible
backreaction, opening the discussion of anisotropy damping and
avoidance of the initial singularity due to quantum 
corrections\cite{HartleHu}.
A semiclassical treatment of the model described by (\ref{act})
with $F(\phi)=1-\xi\phi^2$ and $V(\phi)=\frac{m}{2}\phi^2-
\frac{\Omega}{4}\phi^4$, 
requires the inclusion of
higher orders counterterms to ensure the renormalization of
the theory. These terms are\cite{Birrel,Buchbinder}
\beq
\label{vac}
S_{\rm vac} = \int d^4x\sqrt{-g}\left( 
\alpha_1 R^2 + \alpha_2 R_{ab}R^{ab}
+\alpha_3 R_{abcd}R^{abcd}
+\alpha_4\Box R\right).
\eeq
The quantum 
divergences of the semiclassical
theory can be removed by the renormalization
of the constants $\alpha_{1,2,3,4}$ and the Newtonian constant $G$.
In fact, the full set of quantities affected by the renormalization
includes
the scalar field, its mass $m$ and selfcoupling constant $\Omega$,
the non-minimal coupling constant $\xi$ and  yet a cosmological 
constant. Note that
the last counterterm in (\ref{vac}) does not contribute to the
classical dynamics, since it is merely a total divergence. 

The Weyl tensor $C_{abcd}$ vanishes identically for 
homogeneous and isotropic
spacetimes, and hence, if included in the dynamics,
it 
would affect only the anisotropic case by construction, 
preserving all isotropic
solutions. 
In four
dimensions, we have
\beq
\label{conf}
C_{abcd}C^{abcd}=R_{abcd}R^{abcd}-2R_{ab}R^{ab}+
\frac{1}{3}R^2.
\eeq
Hence, it is possible, in principle, to combine
$\alpha_1$, $\alpha_2$ and $\alpha_3$ in order to have the
counterterm that preserves, by construction,
the isotropic solutions. The resulting action is
\beq
\label{act1}
S+S_{\rm vac}=\int d^4x \sqrt{-g}\left\{\left(
1-\frac{1}{6}\phi^2 
\right)R + \alpha C_{abcd}C^{abcd} - \partial_a\phi\partial^a\phi
-2V(\phi) \right\},
\eeq
where $V(\phi)=\frac{m}{2}\phi^2-\frac{\Omega}{4}\phi^4$ and
$\alpha$ is a parameter typically small when compared to $1/G$.
The action (\ref{act1}) has, by construction, the same
homogeneous isotropic solutions considered in \cite{PRD},
since the term $C_{abcd}C^{abcd}$ and its contributions to
the dynamics vanish in this case.
As to the anisotropic solutions, they are now free of
singularities. The induced counterterm acts a high order
kinetic term for the metric $g_{ab}$, and the Cauchy problem
is well posed everywhere. We can show\cite{GS} that, provided that
$\alpha$ is small, the new term will act only in the region
close to the $F(\phi)=0$ hypersurface. Far from there, the
solutions do not differ much from the ones of (\ref{act}).

A relevant issue is the naturalness of the necessary
adjust in the constants $\alpha_1, \alpha_2$ and $\alpha_3$
in order to have the conformal counterterm. As this constants
may arise from quantum corrections, the only reasonable
hypothesis about them is that they must be small if compared
with $1/G$. This point is now under investigation\cite{GS},
and preliminary results show that, under the only hypothesis
of small $\alpha_1, \alpha_2$ and $\alpha_3$, the singularities
on $F(\phi)=0$ can be eliminated preserving
almost all of the isotropic behavior.

Concluding, in models like (\ref{act}), the passage through the
frontier between $F(\phi)$ positive and negative, which is essential
to the novel interesting cosmological histories described in \cite{PRD},
is classically forbidden as it is unstable at $F=0$ with respect to
anisotropic classical fluctuations. Nevertheless,  
precisely these fluctuations, when considered semiclassically together
their feedback response, render this passage physically meaningful,
enriching dramatically the whole problem and helping us to elucidate
the Physics behind the fantastic   cosmological data available today.

\acknowledgements

The authors acknowledge the financial support from the EEC 
(project HPHA-CT-2000-00015) and FAPESP (Brazil).

\end{document}